\documentclass[sigconf]{acmart}
\AtBeginDocument{%
  \providecommand\BibTeX{{%
    \normalfont B\kern-0.5em{\scshape i\kern-0.25em b}\kern-0.8em\TeX}}}

\setcopyright{none}

\usepackage{graphicx}
\usepackage{caption}
\usepackage{subcaption}
\usepackage[ruled]{algorithm2e}
\usepackage{tcolorbox}

\begin{document}
\setcopyright{none}
\settopmatter{printacmref=false} 
\renewcommand\footnotetextcopyrightpermission[1]{} 
\pagestyle{plain} 
\title{SO\textit{Cluster} - Towards Intent-based Clustering of Stack Overflow Questions using Graph-Based Approach}
\author{Abhishek Kumar, Deep Ghadiyali and Sridhar Chimalakonda}

\affiliation{%
  \institution{\textit{Research in Intelligent Software \& Human Analytics (RISHA) Lab}\\
  Department of Computer Science and Engineering\\
  Indian Institute of Technology Tirupati}
  \city{Tirupati}
  \country{India}
}
\email{{cs17b002, cs17b011, ch}@iittp.ac.in}


\begin{abstract}
Stack Overflow (SO) platform has a huge dataset of questions and answers driven by interactions between users. But the count of unanswered questions is continuously rising. This issue is common across various community Question \& Answering platforms (Q\&A) such as \textit{Yahoo}, \textit{Quora} and so on. Clustering is one of the approaches used by these communities to address this challenge. Specifically, \textit{Intent}-based clustering could be leveraged to answer unanswered questions using other answered questions in the same cluster and can also improve the response time for new questions. It is here, we propose SO\textit{Cluster}, an approach and a tool to cluster SO questions based on \textit{intent} using a graph-based clustering approach.
We selected four datasets of 10k, 20k, 30k \& 40k SO questions without code-snippets or images involved, and performed \textit{intent}-based clustering on them. We have done a preliminary evaluation of our tool by analyzing the resultant clusters using the commonly used metrics of \textit{Silhouette coefficient}, \textit{Calinkski-Harabasz Index}, \& \textit{Davies-Bouldin Index}. We performed clustering for 8 different threshold similarity values and analyzed the intriguing trends reflected by the output clusters through the three evaluation metrics. At 90\% threshold similarity, it shows the best value for the three evaluation metrics on all four datasets. The source code and tool are available for download on Github at: \url{https://github.com/Liveitabhi/SOCluster}, and  the demo can be found here: \url{https://youtu.be/uyn8ie4h3NY}.
\end{abstract}




\maketitle

\section{Introduction}

Stack Overflow (SO) is one of the most successful and commonly used Stack Exchange Network focused on questions related to programming \cite{yelmen2020doc2vec, huang2017expert}. 
SO’s active community attracts information seekers from around the globe harvesting its knowledge-base\cite{asaduzzaman2013answering}.
Despite the rapid growth, there has been a huge rise in the number of unanswered questions on SO \cite{asaduzzaman2013answering}, currently standing at more than 6.4 million\footnote{\url{https://stackoverflow.com/questions?tab=Unanswered}} out of the total 21 million\footnote{\url{https://data.stackexchange.com/}} questions. According to previous studies by Asaduzzaman et al.\cite{asaduzzaman2013answering}, the major reasons for this trend are \textit{Failing to attract an expert member}, \textit{Too short, hard to follow} and \textit{Duplicate question}, altogether accounting for more than 50\% unanswered questions.
Due to the fast-growing user base of SO (currently at around 14 million), the questions count of the platform is rising, thereby causing an increased load on the expert members resulting in many questions not getting required attention.
To maintain the popularity of the platform and to enhance the responsiveness of such services, one can identify similar questions and, thereafter, return the relevant answers from the existing knowledge base of SO platform \cite{john2016graph}. One of the most beneficial solutions to manage this big amount of data is to cluster them automatically according to the similarities\cite{yelmen2020doc2vec}. 
Our key goal is to leverage the clustered questions and answer unanswered questions using other answered questions in that cluster.
\textit{Intent} is one of the key concepts to achieve this goal, which has been used for building dialog systems \cite{haponchyk2018supervised}. Modern search engines go beyond retrieving relevant documents and try to identify the intent of the user query to display relevant results \cite{hashemi2016query}. Inspired from the success of search engines and dialog systems such as \textit{Alexa}, \textit{Cortana} and \textit{Siri} for automatic questions answering, our goal is to cluster questions on SO platform based on \textit{intent} as a way to help programmers. 
The below example explains why focusing on \textit{intent} is important in the context of SO platform.
\begin{itemize}
    \item \textit{Not getting output for merge sort}\footnote{\url{https://bit.ly/3lUU1ew}}
    \item \textit{A bug in merge sort}\footnote{\url{https://bit.ly/3pOxJgX}}
\end{itemize}
In both these questions, users ask about some error/bug present in their implementation of \textit{Merge Sort}. Although the questions look different from each other at the outset, they have similar \textit{intent} and so, the answer of one question might be used to answer the other question if they are in the same cluster.

The idea of clustering of web queries based on \textit{intent} has been extensively explored in the literature \cite{tsur2016identifying, veilumuthu2009intent, kathuria2010classifying, hashemi2016query}. However, web queries can be a collection of random keywords, whereas questions asked to a community for help generally have semantic meaning. For example, \textit{``python lambda function''} can be a web query but to ask on a Q\&A site, one needs to write \textit{``What is Lambda function in python and how to use it?''}. This distinction of web queries from community Q\&A requires a different approach for clustering of questions on a community platform dataset such as the SO.




There have been multiple attempts towards clustering of questions on community platform datasets. Haponchyk et al. \cite{haponchyk2018supervised} have clustered the \textit{Quora} corpus using LSSVM, by training the model on pairwise annotated quora data. Chen et al. \cite{chen2012understanding} used co-training approach to cluster \textit{Yahoo! Answers}. Yelmen \& Duru \cite{yelmen2020doc2vec} clustered SO questions using \textit{doc2vec} vectorizer, however recent BERT models have outperformed \textit{doc2vec} on NLP tasks \cite{mendsaikhan2020identification}. While there is existing work on clustering SO questions\cite{huang2017expert}, they do not focus on \textit{intent}-based clustering, which is the core idea of our approach and tool. There have been various recent works focusing on tasks such as intent-recognition, intent-classification and so on using the latest BERT models\cite{chen2019bert, huggins2021practical}. Huggins et al. confirm that BERT models are quite good in above tasks and achieve 94\% accuracy with minimal training examples \cite{huggins2021practical}. In this work, we leverage Sentence-BERT, a modification of the pretrained BERT network that use siamese and triplet network structures to derive semantically meaningful sentence embeddings \cite{reimers2019sentence}.

Driven by the recent advances in clustering methods based on graphical representations of the relationships among data points\cite{nie2016constrained} and BERT models in intent recognition tasks, we propose
\newline
\noindent\fbox{%
    \parbox{\columnwidth}{%
        \textbf{(i) a graph-based algorithm for intent-based clustering of SO questions and,\\ (ii) SO\textit{Cluster}, a tool based on \textit{Sentence-BERT} vectorizer for creating intent-clusters.}
    }%
}


\section{SO\textit{Cluster} - Design and Development}

\begin{figure}[t]
    \centering
    \includegraphics[width = \linewidth]{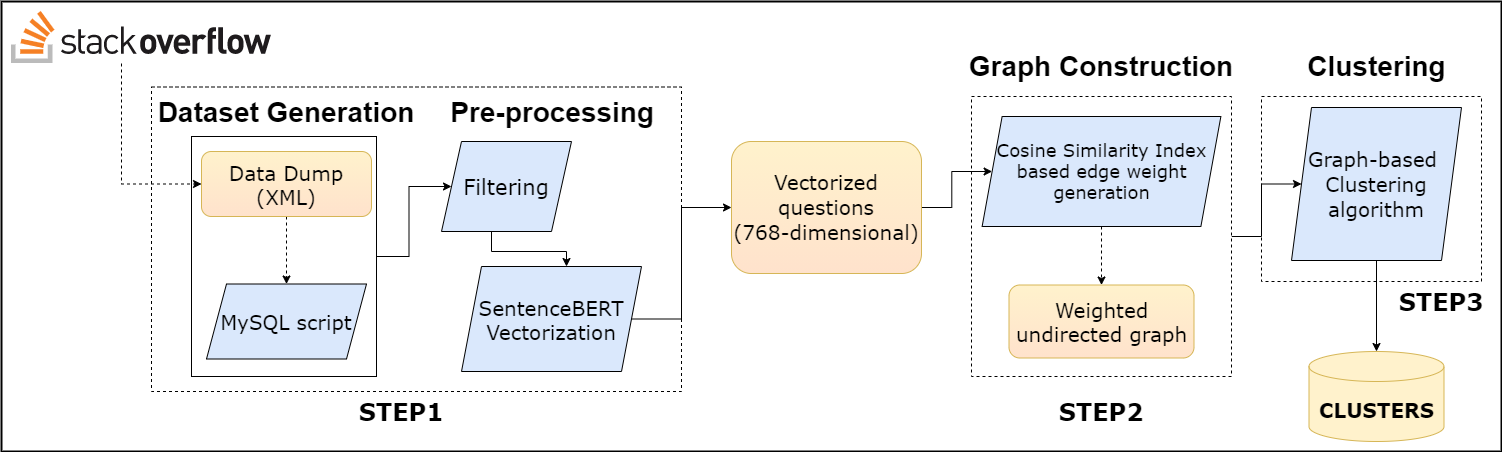}
    \caption{Architecture Diagram of SO\textit{Cluster}}
    \label{fig:arch}
\end{figure}

SO\textit{Cluster} can be divided into three main steps as shown in Figure \ref{fig:arch}. Firstly, we downloaded the SO data dump and processed it using a MySQL script to create well-organized SQL tables. We then filtered and vectorized the dataset using \textit{Sentence-BERT}, which uses recent advances in NLP \& NLU to generate sentence-embeddings\cite{reimers2019sentence}. In the next step, we create a weighted undirected graph where the questions are represented as nodes and edge weight is calculated using the cosine similarity between vector representation of two questions. Lastly, the Clustering algorithm (explained in Algorithm \ref{alg:clustering}) breaks this graph into multiple components by considering only those edges as valid whose weight is greater than the provided threshold similarity. Each of the resulting component in this undirected graph refers to a \textit{cluster}.

\subsection{Step 1: Dataset Generation and Pre-processing}

\subsubsection{Data dump}
We downloaded SO post data from StackExchange data dump archives\footnote{\url{https://archive.org/download/stackexchange}} and obtained the schema for this data from StackExchange \footnote{\url{https://data.stackexchange.com/stackoverflow/query/new}}. 
We then organised these files in MySql database in the form of tables using custom scripts.
\subsubsection{Pre-processing}
We filtered the database by checking it for incorrect or inconsistent data. We ignored questions containing \textit{image}, \textit{table}, \textit{large code snippet} (multiple functions/methods), and so on, as it would require image processing and lexer, parser for processing code snippets. Hence, we focused only on questions with no code snippet involved. In our dataset, the top two tags with most number of questions were \textit{javascript} and \textit{python} with 2081682 and 1528649 questions respectively. Hence, we extracted a subset of the database on \textit{javascript} and \textit{python} tags to perform our experiment, which included both answered as well as unanswered questions. We varied this subset size from 10000 to 40000 questions in four stages.
\subsubsection{Feature Vectorization}
SO\textit{Cluster} uses \textit{Sentence-BERT} for feature vectorization as it considerably improves on BERT model \cite{reimers2019sentence}. 
Attention, Transformers, BERT \& Siamese Network are the four key concepts of Sentence-BERT \cite{reimers2019sentence}. 
The pooling layer in its architecture generates the embedding. It reduced the effort for finding the most similar pair in a collection of 10,000 sentences from 65 hours with BERT/RoBERTa to about 5 seconds, while maintaining the accuracy from BERT \cite{reimers2019sentence}. The number of hidden units in default BERT model architecture is 768 and hence, the output of this vectorization stage is a 768-dimensional vector for each question. BERT models perform good in intent recognition and intent classification tasks as shown by many recent works \cite{chen2019bert, huggins2021practical}. It is in this step that the \textit{Sentence-BERT} vectorizer captures the \textit{intent} of the question and generates the feature vectors.

\subsection{Step 2: Graph Construction}

\subsubsection{Similarity Index}
We used Cosine similarity in our tool to calculate the similarity between two vectors \textit{A} and \textit{B}. It is measured by the cosine of the angle($\varphi$) between two vectors i.e. their inner product space and determines whether they are pointing in roughly the same direction. For our work, we chose Cosine similarity over \textit{Euclidean distance}, \textit{Jaccard coefficient}, \textit{Pearson correlation coefficient} and other metrics because cosine similarity is one of the most popular similarity measure applied to text documents, such as in numerous information retrieval applications and clustering too \cite{huang2008similarity}. Many recent works have used cosine similarity in the field of text document clustering such as Jalal et al. \cite{jalal2021text} to cluster text documents, Rao et al. \cite{rao2021vec2gc} to design an end-to-end term/document clustering pipeline and so on.

\subsubsection{Graph generation}
We created a weighted undirected graph using the feature vectors obtained as nodes and cosine similarity between them as the edge weights. We have used an adjacency matrix representation to store the graph.

\subsection{Clustering}

This step explains the Graph-based Clustering algorithm that SO\textit{Cluster} uses (Algorithm \ref{alg:clustering}). It takes the graph generated in the last step as one of the inputs, along with a \textit{Threshold Similarity Value}. It considers only those edges as valid, whose weight is greater than the threshold value, thereby breaking the graph into multiple components. On finding any unvisited node, it uses \textit{BFS} traversal to search for the connected component and marks the already traversed nodes as \textit{visited}, while edges with lesser weight than the threshold are ignored. Each component returned by the \textit{BFS} procedure here is a \textit{cluster}, and the algorithm returns a set of clusters as output.

\begin{algorithm}
\footnotesize
\SetAlgoLined
\SetKwInOut{KwInput}{input}
\SetKwInOut{KwOutput}{output}
\KwInput{A weighted undirected graph $G$($V,E,W$), Threshold similarity $T_s$ }
\KwOutput{Set of Clusters}

Let Cluster\_Set be a Set\\
$Cluster\_Set = \{\};$\\
\For{each vertex ${k \in V}$}{
    ${Visited[k] = False;}$
}
\For{each vertex ${k \in V}$}{
    \If{${Visited[k] == False}$}{
        $Cluster = BFS(V,k);$\\
        $Cluster\_Set = Cluster\_Set \cup Cluster;$
    }
}
$print$ $Cluster\_Set;$\\
\SetKwProg{myproc}{Procedure}{}{}
\myproc{$BFS(V,k)$}{
    ${Visited[k] = True;}$\\
    Let Q be a Queue\\
    ${Q.insert(k);}$\\
    Let Cluster' be a Set\\
    $Cluster' = \{k\};$\\
    \While{Q is not empty}{
        $s = Q.front();$\\
        $Q.pop();$\\
        \For{each vertex ${p \in V.Adj[s]}$}{
            \If{$Visited[p] == False$ \&\& $W[s][p] \geqslant T_s$}{
                ${Visited[p] = True;}$\\
                ${Q.insert(p);}$\\
                $Cluster' = Cluster' \cup p;$
            }
        }
    }
    \Return{Cluster'};
 }
\caption{Clustering algorithm}
\label{alg:clustering}
\end{algorithm}

\section{Preliminary Evaluation}
To evaluate SO\textit{Cluster}'s performance, we used it to cluster 4 datasets of size 10k, 20k, 30k \& 40k. Since, the clustering algorithm used in SO\textit{Cluster} takes \textit{Threshold Similarity} as one of the inputs (as presented in Algorithm 1), we performed the experiment by changing this parameter over eight different values : 0.5, 0.6, 0.65, 0.7, 0.75, 0.8, 0.85 \& 0.9, for each dataset and observed the results.

To evaluate these outputs, we used three commonly used clustering evaluation metrics in the literature, which are used for clustering performance evaluation when ground truth labels are unavailable \cite{unlu2019estimating}. The \textit{Silhouette Coefficient} estimates the similarity of an object to its cluster compared to the next-nearest cluster, \textit{Calinski-Harabasz Index} score is defined as the ratio of the sum of between-clusters dispersion and inter-cluster dispersion for all clusters and the \textit{Davies-Bouldin Index} evaluates the cluster using quantities and features inherent to the dataset. We calculated these three metrics for clusters obtained for varied dataset sizes and threshold values. 

\section{Results}

\begin{figure}[htbp]
  \centering
  \begin{subfigure}{.5\columnwidth}
    \centering
    \includegraphics[width=\linewidth]{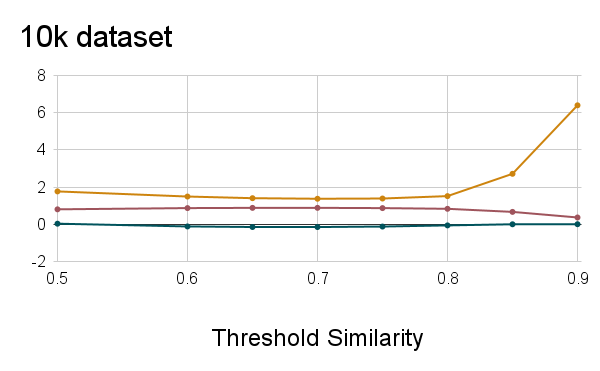}
    \caption{Dataset : 10000}
  \end{subfigure}%
  \hfill
  \begin{subfigure}{.5\columnwidth}
    \centering
    \includegraphics[width=\linewidth]{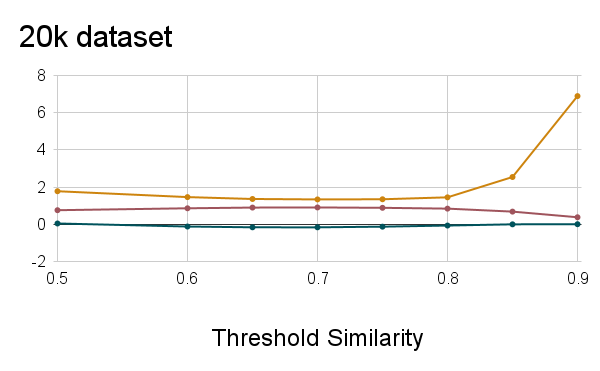}
    \caption{Dataset : 20000}
  \end{subfigure}%
  \hfill
  \centering
  \begin{subfigure}{.5\columnwidth}
    \centering
    \includegraphics[width=\linewidth]{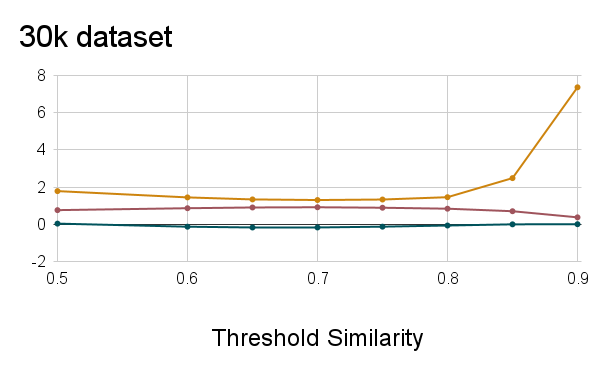}
    \caption{Dataset : 30000}
  \end{subfigure}%
  \hfill
  \begin{subfigure}{.5\columnwidth}
    \centering
    \includegraphics[width=\linewidth]{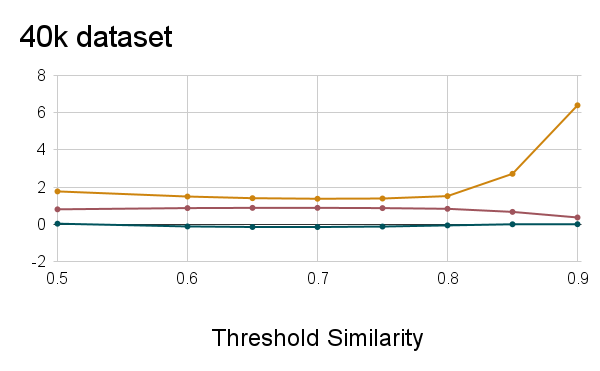}
    \caption{Dataset : 40000}
  \end{subfigure}%
  \hfill
  \begin{subfigure}{\columnwidth}
    \centering
    \includegraphics[width=\linewidth]{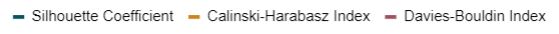}
  \end{subfigure}%
  \hfill
  \caption{Results of evaluation metrics using SO\textit{Cluster}}
\label{fig:eval}
\end{figure}

The resultant clusters obtained across the experiments were of different sizes. The spread of the cluster sizes was narrow for both small and big thresholds but it was wide for medium threshold values such as 0.75 \& 0.80 . For small threshold, a big cluster was formed in all cases. For bigger thresholds, maximum clusters had a single element. But medium thresholds such as 0.75 \& 0.80 were optimum for the clusters to spread widely.
The graphs in Figure \ref{fig:eval} summarize the trend of the three evaluation metrics across all experiments. We observe that the results of evaluation are not much affected by change in the dataset size, but we note that our sample is from 10,000 to 40,000 whereas the number of questions and unanswered questions are in the order of millions on SO. 

However, for a fixed dataset, increase in threshold value beyond 0.7 shows an improvement in the evaluation metric values. The reason behind this might be that threshold similarity of 0.80, 0.85, 0.90 and above impose a very strict similarity condition clustering only near-identical questions together.

\begin{figure}[htbp]
    \centering
    \includegraphics[width = \linewidth]{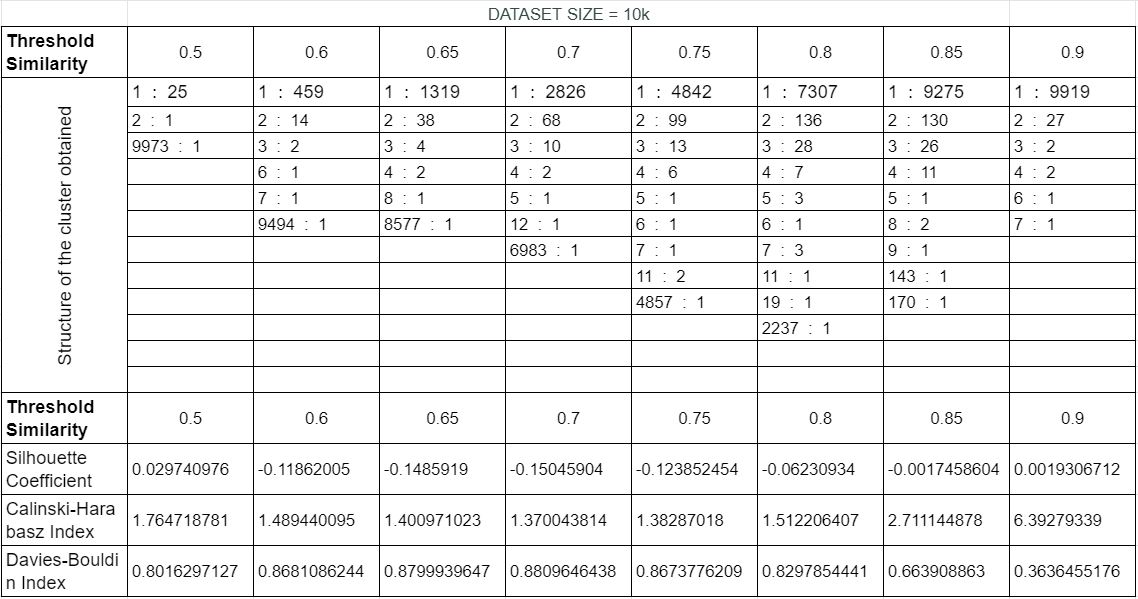}
    \caption{Distribution of Clusters and evaluation metrics for 10k dataset using SO\textit{Cluster} - Format X : Y where X = cluster size and Y = cluster count}
    \label{fig:res_10k}
\end{figure}
Table in Figure \ref{fig:res_10k} shows the result for dataset of 10k questions.
The detailed results of clustering of different dataset sizes with varying threshold values can be found at: \href{https://docs.google.com/spreadsheets/d/1W3BAtn6g3-76W0ZQXA7SWixJE_MvxdF1PjLBSwkW6uE/edit?usp=sharing}{\underline{results}}.
 
A closer look over the resultant clusters reveals interesting insights. The spread of the clusters over different sizes tells how "good" a cluster is for our intended goal of answering unanswered questions. More evenly spread cluster means a better chance of the unanswered questions falling in group with an answered question.
If there are too many specific and different questions which lead to small size clusters, then those clusters cannot be efficiently used to answer most of the unanswered question.
Changing the dataset to include questions which are more related to each other (belonging to same domain or language) is a solution for this. In our case, this trend can be seen for very high threshold similarity values. Also, Large clusters can be troublesome too, if they group most of the questions together, clustering won't be able to serve its purpose. So, the aim should be to achieve clusters with a more even spread over different sizes. It means the dataset has potential for handling some of its unanswered questions by itself. Thus, developers can also use SO\textit{Cluster}'s results to evaluate their dataset and then, can work on those datasets towards automatic answering.

\section{Discussion and Limitations}
We aimed at \textit{intent}-based clustering of questions on SO platform with the ultimate goal of answering millions of unanswered questions using answered questions. However, we observed several challenges during our research. Firstly, there are many singleton clusters i.e. clusters with only one entity. We had varied singleton cluster ratio (SCRs), which is the ratio of singleton clusters to total output clusters for different dataset sizes. When the dataset size is too small ($<$20 questions), the SCR is too high ($\sim$1, which means mostly singleton cluster) because with few questions, the odds of two or more questions being similar is low. With the increase in dataset size, the SCR remained close to 1 (decreasing very slowly) in our experiment. But, as we further increase the size of the dataset (>40,000 questions), we believe that SCR might come down as chances of two or more questions getting similar increases with wide spread across the clusters. However, using a serial \textit{BFS} traversal algorithm becomes a bottleneck with respect to time. 


For the same dataset, when we increased the threshold, the spread of clusters widened, and then it reached to a maxima, and then again narrowed down for higher values of the threshold. The results also indicate that the chances of two questions to fall in the same cluster decreases as the value of threshold increases.
Ideal value of the threshold according to our experiment lies in between 0.8 to 0.9. Threshold values greater than 0.9 will be too strict for clustering the questions. Also, lower values of thresholds (<0.5) resulted in clustering of unsimilar questions.

An inherent limitation of the tool is not to consider questions having code snippets (multiple functions/methods) or images in order to avoid the overhead of processing images or code snippets. Wu et al.\cite{wu2019developers} observed that 75\% of the answers on SO have at least one source code snippet attached. We believe that the tool can be extended and integrated with code vectorization, code summarization or image processing techniques. Also, adding a UI or plugin for SO\textit{Cluster} can make it more user-friendly and easy-to-use.
Finally, the algorithm and metrics we used for clustering could be further improved, along with validation from user studies. 

\section{Related Work}

Clustering of large datasets from Q\&A sites has attracted the attention of many researchers. Existing literature has a lot of applications in Question \& Answering \cite{john2016graph}, Dialog System Applications \cite{haponchyk2018supervised} and other domains. Many earlier works have contributed to this through various algorithms and models.

Chen et al. \cite{chen2012understanding} have used a semi-supervised learning technique called \textit{co-training} approach on \textit{Yahoo! Answers} corpus to understand user intent by classifying them into subjective, objective and social. But using these predefined cluster-labels has its limitation of missing out on some important class. Nie et al. \cite{nie2016constrained} proposed the Constrained Laplacian Rank Algorithm for Graph-Based Clustering. It takes the data graph as input and allows it to be adjusted as part of the clustering procedure so that the quality of the resulting clustering is not affected by low quality input data graph. This CLR-algorithm has been applied and tested against SO dataset  by Huang et al. \cite{huang2017expert}, where they have used a term frequency based representation of posts. However, their term-frequency based representation does not capture the essence of the \textit{Intent} of the questions. 

Yelmen \& Duru \cite{yelmen2020doc2vec} used K-Means++, K-Mediods \& Gaussian Mixture to perform clustering of SO posts. However, they have used doc2vec word embedding method and recent study shows that BERT models have outperformed doc2vec on cybersecurity-related NLP tasks  \cite{mendsaikhan2020identification}. Our approach uses SentenceBERT vectorizer which has been observed to perform better than BERT \cite{reimers2019sentence}.
Chen \& Zing \cite{chen2016mining} have mined technological landscapes from SO by creating \textit{community-clusters}. Villanes et al. \cite{villanes2017software} have clustered questions using LDA algorithm on only Android testing domain, but not based on \textit{intent}. Unlike the machine learning based approach in Beyer et al. \cite{beyer2018automatically} and LDA-based approach in Venigalla et al. \cite{venigalla2019sotagger} for intent-based classification of SO posts, SO\textit{Cluster} uses graph-based approach for intent-based clustering.

\section{Conclusion and Future Work}
Stack Overflow is an important Q\&A based knowledge-sharing community for programmers and developers. Clustering the questions on SO based on \textit{intent} can be used to answer millions of unanswered questions on the platform. In this paper, we proposed SO\textit{Cluster}, an approach and a tool that clusters SO Q\&A dataset using graph-based clustering approach. For the demonstration, random questions from SO posts with varying size (10,000 to 40,000) were provided as an input to the tool and clustered for 8 different threshold values of similarity index. \textit{Intent-clusters} were obtained as an output containing both answered and unanswered questions. The spread of the sizes of these clusters was narrow for too low or high threshold similarities and optimum for medium thresholds. We demonstrated through the SO\textit{Cluster} tool that a graph-based approach for intent-based clustering has potential to answer unanswered questions on the SO platform. 

Our future goal is to improve the evaluation for optimum cluster-size distribution. We plan to improve the vectorization of the questions by training the model and also plan to include the intent of image or code snippets and cluster the posts which include them. We also plan to consider forming cluster on larger dataset and to implement a parallel version of the BFS algorithm to reduce the run-time of the clustering process. 
\newpage




\balance

\bibliographystyle{ACM-Reference-Format}
\bibliography{socluster}


\end{document}